\documentclass[10pt,letterpaper]{article}
\usepackage{arydshln}
\usepackage{authblk}
\usepackage[utf8]{inputenc}
\usepackage[backend=biber,style=numeric-comp,sortcites=true,minnames=6,maxbibnames=7,maxcitenames=7,giveninits=true,terseinits=true,sorting=none]{biblatex}
\usepackage[singlelinecheck=false]{caption}
\usepackage{comment}
\usepackage{fancyhdr}
\usepackage[a4paper]{geometry}
\usepackage{graphicx}
\graphicspath{{Figures/}}
\usepackage[pagewise]{lineno}
\usepackage{multirow}
\usepackage{parskip}
\usepackage{subcaption}
\usepackage{todonotes}
\usepackage{ulem}
\usepackage{verbatim}
\usepackage{wasysym}

\bibliography{bibliography.bib}

\DeclareFieldFormat[book, article, thesis, inproceedings]{title}{#1}
\DeclareFieldFormat{journaltitle}{#1}
\DeclareFieldFormat{booktitle}{#1}
\DeclareFieldFormat{postnote}{\mkcomprange[{\mkpageprefix[pagination]}]{#1}}
\DeclareFieldFormat{pages}{\mkcomprange{#1}}

\DeclareNameAlias{default}{family-given}

\renewbibmacro{in:}{\ifentrytype{article} {} {\printtext{\bibstring{in} \intitlepunct}}}

\renewbibmacro*{journal+issuetitle}{%
  \usebibmacro{journal}%
  \setunit*{\addperiod\addspace}%
  \iffieldundef{series}
    {}
    {\newunit
     \printfield{series}%
     \setunit{\addspace}}%
  \setunit{\addperiod\addspace}%
  \usebibmacro{issue+date}%
  \setunit{\addcolon\space}%
  \usebibmacro{issue}%
  \setunit{\addsemicolon\space}%
  \usebibmacro{volume+number+eid}%
  \newunit}
  
\DeclareFieldFormat[article,periodical]{number}{\mkbibparens{#1}}
\renewbibmacro*{volume+number+eid}{%
  \printfield{volume}%
  \printfield{number}%
  \setunit{\addcomma\space}%
  \printfield{eid}}

\renewbibmacro*{issue+date}{%
  \iffieldundef{issue}
    {\usebibmacro{date}}
    {\printfield{issue}%
     \setunit*{\addspace}%
     \usebibmacro{date}}%
  \newunit}

\makeatletter
\renewcommand{\maketitle}{\bgroup\setlength{\parindent}{0pt}
\begin{flushleft}
  \Large{\textbf{\@title}}
  
  \vspace{2ex}
  \textit{\normalsize{\@author}}
\end{flushleft}
\egroup
}

\makeatother

\title{An innovative dose rate independent 2D Ce-doped YAG scintillating dosimetry system for time resolved beam monitoring in ultra-high dose rate electron “FLASH” radiation therapy.}
\author[a,b$\dagger$]{\underline{Vanreusel Verdi}}
\author[c]{Heinrich Sophie}
\author[d]{De Kerf Thomas}
\author[e]{Leblans Paul}
\author[e]{Vandenbroucke Dirk}
\author[d]{Vanlanduit Steve}
\author[b,f]{Verellen Dirk}
\author[b,f]{Gasparini Alessia}
\author[a]{de Freitas Nascimento Luana}
\affil[a]{Research in Dosimetric Applications, SCK CEN, Boeretang 200, 2400 Mol, Belgium}
\affil[b]{AReRO, University of Antwerp, Universiteitsplein 1, 2610 Wilrijk, Belgium}
\affil[c]{Institut Curie, Inserm U 1021-CNRS UMR 3347, University Paris-Saclay, PSL Research University, Orsay, France}
\affil[d]{InViLab, University of Antwerp, Groenenborgerlaan 171, 2020 Antwerpen, Belgium}
\affil[e]{Innovation office, Agfa N.V., Septestraat 27, 2640 Mortsel, Belgium}
\affil[f]{Department of Medical Physics, Radiotherapy, Iridium Netwerk, Oosterveldlaan 22, 2610 Wilrijk, Belgium}

\begin{document}
\maketitle
\vspace{5ex}
\section*{Abstract} 
Preclinical studies have shown that radiotherapy treatments benefit from ultra-high dose rate (UHDR) irradiations. These trigger the FLASH-effect, resulting in a strong decrease in normal tissue toxicity with conserved tumor control probability, compared with conventional irradiations.
However, the beam parameters to trigger the FLASH-effect and the radiobiological mechanisms behind it remain to be elucidated. A limiting factor in many studies is the lack of accurate real time dosimetry.  
In this work an innovative solution for 2D real time dosimetry in UHDR electron beams is presented and characterized.
\par
The in-house developed ImageDosis system consists of a scientific camera, with high temporal resolution, and a coating, containing 12\% of Y$_3$Al$_5$O$_{12}$:Ce$^{3+}$ as scintillating material. Reference dosimetry was performed by means of radiochromic film, and a (C$_{38}$H$_{34}$P$_2$)MnBr$_4$ point scintillator was used to validate the pulse discrimination properties of the ImageDosis system.
Irradiations were performed in two centers (Antwerp and Orsay), with an ElectronFlash accelerator. The ImageDosis system was tested with varying number of pulses, pulse length, pulse repetition frequency (PRF) and energy. In addition, its temporal resolution and 2D properties were investigated.
\par

\par
The ImageDosis system showed negligeable dose, dose rate and energy dependence for doses up to 13 Gy and average dose rates up to 140 Gy/s, with a dose per pulse up to 2 Gy and a PRF up to 300 Hz. It showed capable of discriminating and measuring the dose of individual pulses and promising 2D characteristics that need further optimization.

\vfill

\flushleft{$^\dagger$\textit{Corresponding author:} verdi.vanreusel@sckcen.be \newline \textit{Address:} Boeretang 200, 2400 Mol, Belgium}\\\vspace{2ex}
\textit{\textbf{Keywords:} FLASH radiotherapy; UHDR dosimetry; real time dosimetry; 2D dosimetry; Scintillating sheets} 
\thispagestyle{fancy}
\fancyhf{}
\renewcommand{\headrulewidth}{0pt}
\fancyfoot[L]{\textbf{\large Abbreviations} \vspace*{0.5ex} \newline
                FWHM: Full width at half maximum\newline
                PRF: Pulse repetition frequency\newline
                ROI: Region of interest \newline
                RT: Radiotherapy \newline
                UHDR: Ultra-high dose rate\newline
                YAG: yttrium aluminum garnet}

\newpage
\setcounter{page}{1}
\section{Introduction}
Radiotherapy (RT) is one of the major strategies for treating cancer, however, its efficacy is limited due to the normal tissue toxicity associated with ionizing radiation. Even though delivery techniques have improved and enabled the prescription of higher doses in a single treatment or fewer fractions~\cite{verellen2007innovations, abshire2018evolution}, severe side effects are still observed with many pathologies, especially in the case of radioresistant tumors and when target volumes intersect with organs at risk~\cite{lindberg2017oa24, salloum2019late}. The discovery of the FLASH-effect~\cite{favaudon2014ultrahigh} was, therefore, positively received, and is expected to play an important role in the future of radiotherapy treatments~\cite{wilson2020ultra}. The FLASH-effect is a biological effect in which a strong reduction of adverse effects is observed when using ultra-high dose rates (UHDR) to deliver the treatment~\cite{bourhis2019clinical,wilson2020ultra}. The radiobiological mechanisms underlying this effect are yet to be elucidated. Several (concurrent) hypotheses are being considered, yet remain to be confirmed~\cite{marcu2021translational,friedl2022radiobiology,borghini2022flash,taylor2022roadmap}. Reproducible preclinical research using accurate real time dosimetry is of paramount importance to understand the radiobiological mechanisms of the FLASH-effect and determine the threshold irradiation parameters to obtain the FLASH-effect. An important aspect for this latter is the dose rate definition. In pulsed electron beams, the average dose rate and instantaneous dose rate (dose rate within the pulse) are considered relevant. Equally important are the properties that define these dose rates, being the dose per pulse, pulse length and pulse repetition frequency (PRF)~\cite{vozenin2019biological,vozenin2020all}. In pencil beam scanning proton beams the dose is delivered sequentially by multiple spots, and the dose rate in a point is defined by all the spots contributing to the dose deposition in that point. Therefore, the dose rate definition is not straight-forward and multiple definitions exist~\cite{folkerts2020framework,van2020bringing,deffet2023definition}.   

UHDR dosimetry is a basic requirement to translate FLASH-RT to the clinic, to assure a precise measurement and to monitor the extremely high dose rates involved in this emerging radiotherapy technique. However, achieving accurate dosimetry in FLASH radiotherapy remains a challenge, primarily due to limitations in real time measuring devices and the complex nature of delivering UHDRs. The dosimeters typically used in conventional radiotherapy tend to saturate~\cite{di2020flash}. Ionization chambers, for example, are the standard in conventional radiotherapy, but are subject to high ion recombination effects in UHDR beams. The relevant correction factors rise up to more than 70\%, and strongly reduce the reliability of the dose measurement, especially for pulsed electron beams~\cite{mcmanus2020challenge, kranzer2021ion, romano2022ultra, kranzer2022charge}. International work groups (AAPM TG359, UHDPulse~\cite{schuller2020european}, etc.) on the topic have resulted in the investigation and development of many dosimeters for UHDR irradiations, ranging from passive point dosimeters such as thermoluminescent dosimeters and alanine EPR dosimetry~\cite{bourgouin2022absorbed}, passive 2D dosimeters such as Gafchromic film~\cite{jorge2022design} and optically stimulated luminescence sheets~\cite{vanreusel2023optically}, and real time point dosimeters such as the FlashDiamond (PTW, Germany)~\cite{marinelli2022design,verona2022application,marinelli2023diamond}; the small portable graphite calorimeter~\cite{bourgouin2022probe}; the ultra thin parallel plate ionization chamber~\cite{gomez2022development}; point scintillators~\cite{vanreusel2022point}; and water~\cite{bourgouin2023ptb} and graphite calorimetry~\cite{lourencco2023absolute}. Despite the positive results obtained with these dosimeters, none of them have the capability of 2D real time dose assessment. Recently, three studies investigated a setup using cameras with scintillating sheets or Cherenkov radiation for 2D real time dosimetry and beam monitoring. Promising results were obtained in both particle beams and electron beams with converted clinical linacs~\cite{rahman2020characterization,rahman2021spatial, levin2023scintillator}. 

In this study we propose an innovative solution for 2D real time dosimetry in UHDR electron beams, where a scintillating material is coated on a 2D sheet. However, in contrast with Rahman et. al.~\cite{rahman2021spatial} where the scintillating material is $Gd_2O_2S:Tb$, with a scintillating decay time of $\sim$ 600 µs, dose rate independence up to 300 Gy/s and dose per pulse measurement up to 60 Hz irradiations, we propose a solution based on a Cerium doped yttrium aluminum garnet (YAG) (Y$_3$Al$_5$O$_{12}$:Ce$^{3+}$) scintillating sheet, with a scintillating decay time of $\sim$ 70 ns~\cite{xia20173+}.

The use of YAG as scintillator material in the medical field has been studied since the 1990s. Its transparency, short lifetime (ns range) and the absence of afterglow make it a useful material for PET detectors~\cite{xia20173+}. These characteristics also suggest that YAG can be used to measure individual pulses even for high frequency radiotherapy irradiations. It has already been studied as point scintillator and has shown a high dose-response linearity~\cite{jia2021tapered} and the capability to measure individual pulses in conventional photon beams with pulse repetition frequencies up to 400 Hz~\cite{chen2019investigation}.

The objective of this paper is to provide a comprehensive dosimetric characterization of the real time two-dimensional (2D) dosimetry system for UHDR electron “FLASH” radiation therapy. The response of the YAG-based system with dose; its average dose rate, dose per pulse, PRF and energy (in)dependence, its capability to measure individual pulses and its 2D characteristics are presented and discussed.

\section{Materials and Methods}

\subsubsection*{ImageDosis 2D real time system}
The ImageDosis system is an in-house created system, consisting of a scintillating coating and a scientific camera~\cite{nascimento2020two,nascimento2021real}. The coating was composed of a binder, mixed with 12\% of YAG:Ce (Y$_3$Al$_5$O$_{12}$:Ce$^{3+}$) scintillating material~\cite{yan2021dosimeter}. The emission wavelength of YAG:Ce peaks at 551 nm (green) and the decay time is 70 ns~\cite{xia20173+}. The coating has a size of 20 cm x 25 cm and a thickness of 0.1 mm. 
\par
To record the scintillating emission, the complementary metal-oxide-semiconductor (CMOS) C-BLUE ONE camera (First Light Imaging, France) with a Basler Lens C125 5M (Basler AG, Germany) was used. The camera was operated with the First Light Vision software. The frame rate was chosen such that it was as close as possible to 500 frames per second. The maximal exposure time corresponding with this frame rate (i.e. the inverse of the frame rate minus the processing time of the camera) was chosen. No triggering was used due to the unavailability of the trigger connection between the camera and the ElectronFlash. Recordings were stored as image stacks in tiff format (unsigned 16 bit integers). The most relevant features and settings of the camera are listed in Table~\ref{tab: camera settings} (supplementary section).
\subsubsection*{Point Scintillator real time system}
An in-house made (C$_{38}$H$_{34}$P$_2$)MnBr$_4$ point scintillator with PMMA optical fiber was attached to the applicator as secondary dosimeter for pulse discrimination. The optical fiber system was optimized for UHDR electron irradiations and is described in~\cite{nascimento2014medical,vanreusel2022point}. The hardware has been updated since our previous work~\cite{vanreusel2022point}, with a faster data acquisition card and a new photomultiplier tube, allowing higher sampling rates. A sampling rate of 1 kHz was used. 
\subsubsection*{Radiochromic films}
Pieces of EBT XD or EBT-3 (Ashland, USA) of at least 2 x 2 cm$^2$ were used as reference dosimeter to account for possible output variations from the UHDR electron beam. These films were scanned 2-3 days post irradiation with an Epson 11000XL scanner (Epson, Japan) at 150 dpi resolution.

\subsection{Irradiations}
The ImageDosis system was investigated for real time 2D UHDR dosimetry in two measurement campaigns. It was first characterized and calibrated in Antwerp (Antwerp University Hospital, Belgium) after which it was validated in Orsay (Institut Curie, France). Both facilities are equipped with an ElectronFlash linac (S.I.T., Italy), which allows systematic variation of the beam characteristics of a pulsed electron beam. These include the dose-rate modality (conventional dose rate and UHDR), energy, pulse repetition frequency (PRF), number of pulses, and pulse length. 
\subsubsection*{Characterization}
\begin{figure}[!ht]
    \centering
    \includegraphics[width = .45\textwidth]{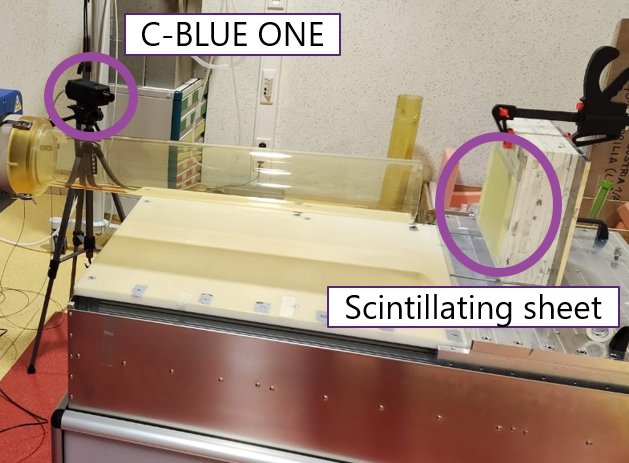}  
    \caption{The setup used. The components of the ImageDosis system (the camera and scintillating sheet) are highlighted in the purple circles.}
    \label{fig: setup}
\end{figure}

In Antwerp the dose response of the 2D scintillating system was characterized in UHDR modality, for an electron beam with nominal energy of 9 MeV.
The irradiations were performed using an applicator with a nominal diameter of 12 cm $\pm$ 0.1 cm with length 91.2 cm $\pm$ 0.1 cm. The scintillating sheet was positioned at a distance of 105.0 cm $\pm$ 0.1 cm from the linac-applicator connection. It was taped against a RW3 phantom of at least 5 cm water equivalent thickness. The camera was positioned at a distance of 72 $\pm$ 5 cm from the central axis, at the height of the linac-applicator connection, focused at the center of the sheet as shown in Figure~\ref{fig: setup}.
\newline
The dose response stability was investigated for variation of the dose; pulse length; PRF; and energy. The dose was varied via the number of pulses (1-23), while keeping the pulse length and PRF constant at 1 µs and 50 Hz, respectively. The pulse length was varied between 0.5 and 4 µs, while keeping the PRF and dose constant at 50 Hz and 4.5 $\pm$ 0.8 Gy, respectively. The PRF was varied between 1 and 245 Hz, while keeping the pulse length and dose constant at 1 µs and 5.3 $\pm$ 0.1 Gy. The energy was varied between nominal values of 7 and 9 MeV, while keeping the pulse length and PRF fixed at 1 µs and 50 Hz, respectively. For the 7 MeV beam 56 pulses were delivered, resulting in a dose of 10.3 $\pm$ 0.4 Gy and for the 9 MeV beam, 9 pulses were delivered, resulting in a dose of 5.4 $\pm$ 0.2 Gy.  
\newline
Two additional field size measurements were performed for applicators with nominal diameters of 1.8 cm and 10 cm, and lengths of 21.5 $\pm$ 0.1 cm and 73.5 $\pm$ 0.1 cm, respectively. For the 1.8 cm applicator, the scintillating sheet was positioned at a distance of 35.7 $\pm$ 0.1 cm from the linac-applicator connection, and the camera was placed at a distance of 30 cm from the central axis, at the position of the linac-applicator connection, focused at the center of the sheet. For the 10 cm applicator, these respective distances were 88.4  $\pm$ 0.1 cm and 72 cm.
\subsubsection*{validation}
In Orsay, alike experiments were performed to validate the results obtained in Antwerp and investigate the transferability of the ImageDosis system. 
The setup was reproduced as good as possible. The applicator with a nominal diameter of 12 cm $\pm$ 0.1 cm was used and the positioning of the camera and sheets were the same as for the experiment in Antwerp, within the positioning uncertainty. A local PMMA phantom, with at least 5 cm water equivalent thickness, was used as opposed to the RW3 phantom in Antwerp.
\newline
The dose response stability was validated for variation of the dose; pulse length; PRF; and energy. Except for the energy variation, a nominal beam energy of 5 MeV was used for all irradiations. The dose was varied via the number of pulses (1-30), while keeping the pulse length and PRF constant at 1 µs and 50 Hz, respectively. The pulse length was varied between 0.69 and 3.98 µs, while keeping the PRF and number of pulses constant at 50 Hz and 7 pulses, respectively. The PRF was varied between 50 and 300 Hz, while keeping the pulse length and dose constant at 1 µs and 7.1 $\pm$ 0.2 Gy. The energy was varied between nominal values of 5 and 7 MeV, while keeping the pulse length and PRF fixed at 1 µs and 50 Hz, respectively. For the 5 MeV beam 30 pulses were delivered, resulting in a dose of 11.6 $\pm$ 0.4 Gy and for the 7 MeV beam, 20 pulses were delivered, resulting in a dose of 15.2 $\pm$ 1.1 Gy.
\par
The irradiation settings of each individual irradiation from both measurement campaigns, can be found in Table~\ref{tab: Beam parameters} (supplementary section). All irradiations were performed in triplet, and the error bars denote one standard deviation. All lights were turned off and light sources such as radiation indicators were blocked during the irradiations to reduce background noise.
\par
A piece of radiochromic film was placed in the field, slightly outside the center to allow the selection of a non-overlapping region of interest (ROI) for signal extraction (Figure~\ref{fig: ROIDef}). For the relative measurements investigating the 2D performance of the ImageDosis system, a large piece of radiochromic film was used to encompass the entire field. In Antwerp, a piece of radiochromic film was used only for one measurement per irradiation settings. 
\par
Next to radiochromic film, also the point scintillator, taped to the applicator, was used in Orsay for investigation of the real time characteristics of the ImageDosis system. 

\subsection{processing}
\begin{figure}[!ht]
    \centering
    \begin{subfigure}{.45\textwidth}
        \includegraphics[width = \textwidth]{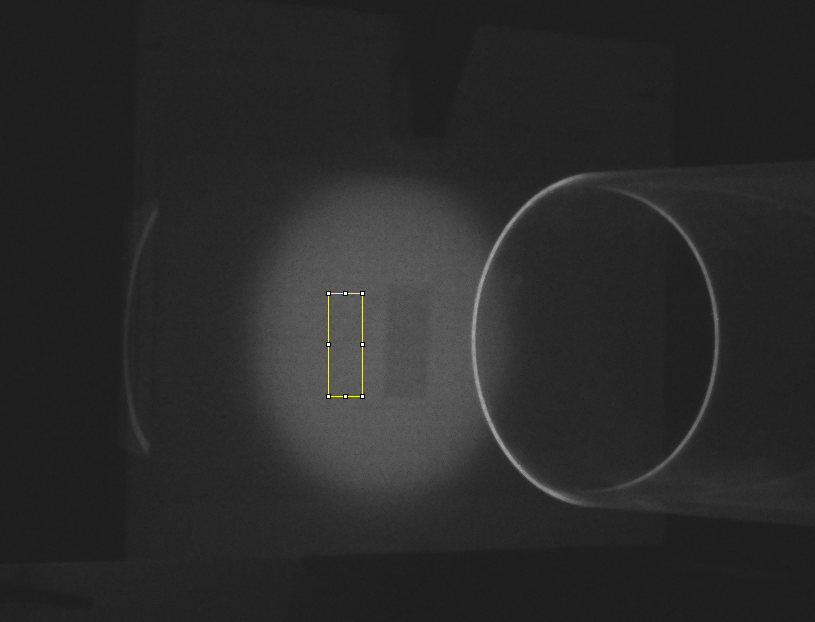}
        \caption{\centering}
        \label{fig: ROIDef}
    \end{subfigure}
    \begin{subfigure}{.45\textwidth}
        \includegraphics[width = \textwidth]{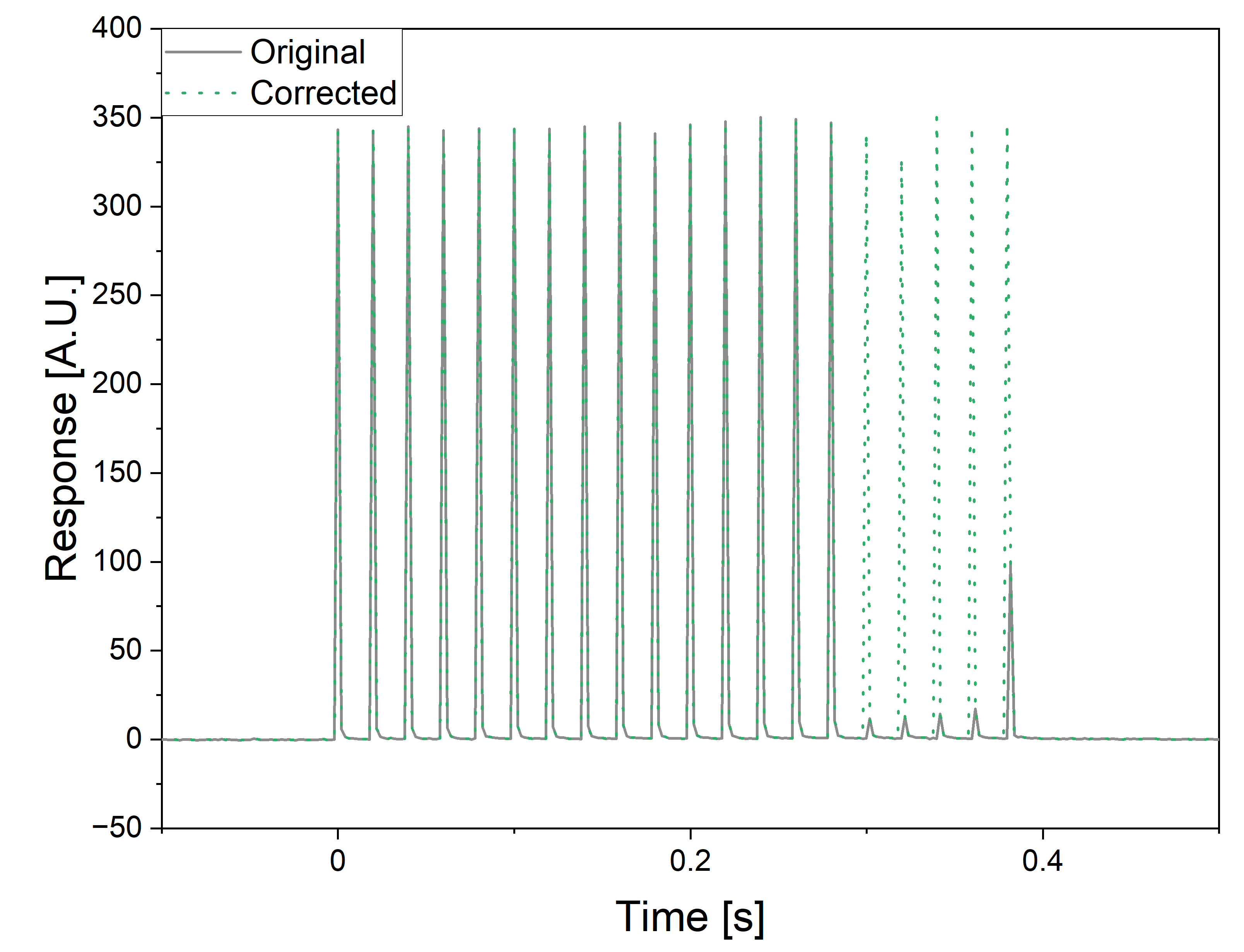}
        \caption{\centering}
        \label{fig:TimeTraceCorrection}
    \end{subfigure}
    \caption{(a) The first pulse with ROI definition (yellow box) inside the field, but not overlapping with the piece of radiochromic film (darker rectangle). (b) ImageDosis time trace before and after correction for missing pulses based on the point scintillating system.}
\end{figure}

The image stacks from the ImageDosis system consist of frames that are either uniform black or show an elongated circular bright spot with a dark square, where the film was positioned, and the outline of the applicator, as shown in Figure~\ref{fig: ROIDef}. The former frames are the ones where no pulse is being delivered during the exposure time, while the latter are the frames where a pulse is being delivered during the exposure time. 
The image stacks were truncated and denoised in Fiji~\cite{schindelin2012fiji}, after which a global background subtraction was performed as described hereafter. For the truncation, a Fiji macro automatically detects the frames corresponding to the first and last pulses of the irradiation. It selects the frames from 100 frames prior to the first pulse to 100 frames succeeding the last pulse. Then, these frames are denoised with a median filter with radius of 2 pixels and threshold value of 5. The first and last 25 frames are averaged to create a background image. The background is defined as the average gray value within a ROI on that background image. This ROI is manually chosen in the frame of the first pulse, such that it falls well within the field and does not overlap with the radiochromic film. An example is shown in Figure~\ref{fig: ROIDef}. Its location is kept constant within all irradiations of an experiment. The response corresponding to a frame is defined as the average gray value of the ROI, subtracted by the background value. This response is calculated for every frame to obtain a time trace of the irradiation. 
\par
Since no triggering was used, pulses that were (partially) in the dead time (being 2.8\% of the exposure time) of the camera could not be detected or were strongly underestimated. The response of these pulses in the time trace was corrected based on the point scintillator data, when available, using the following protocol. First, the point scintillator time trace was resampled to match the ImageDosis time trace, after which it was normalized to its maximum signal. A correction was performed if the amplitude of a pulse from the ImageDosis time trace (normalized to its maximum signal) was less than 80\% of the amplitude of the matching pulse from the point scintillator. The response, extracted from the ROI, of the pulse in the ImageDosis time trace was then replaced by the corresponding response from the point scintillator time trace, multiplied by the maximal signal from the ImageDosis time trace. An example of an ImageDosis time trace before and after correction can be seen in Figure~\ref{fig:TimeTraceCorrection}. 
\par
If no point scintillator data were available, time traces with missing pulses or pulses with reduced amplitude were excluded.
\par
The corrected ImageDosis time trace was integrated and normalized by the radiochromic film dose to obtain the response per Gy. The stability of the resulting data points is investigated to assess the (in)dependence of the ImageDosis system under various irradiation conditions.
\par
For the comparison of the profiles, the penumbra and the field size were defined as the average (left and right) 20\%-80\% width of the dose profile, and the full width at half maximum (FWHM), respectively.

\section{Results}
\subsection{Characterization}
\begin{figure}[!ht]
    \centering
    \includegraphics[width = .5\textwidth]{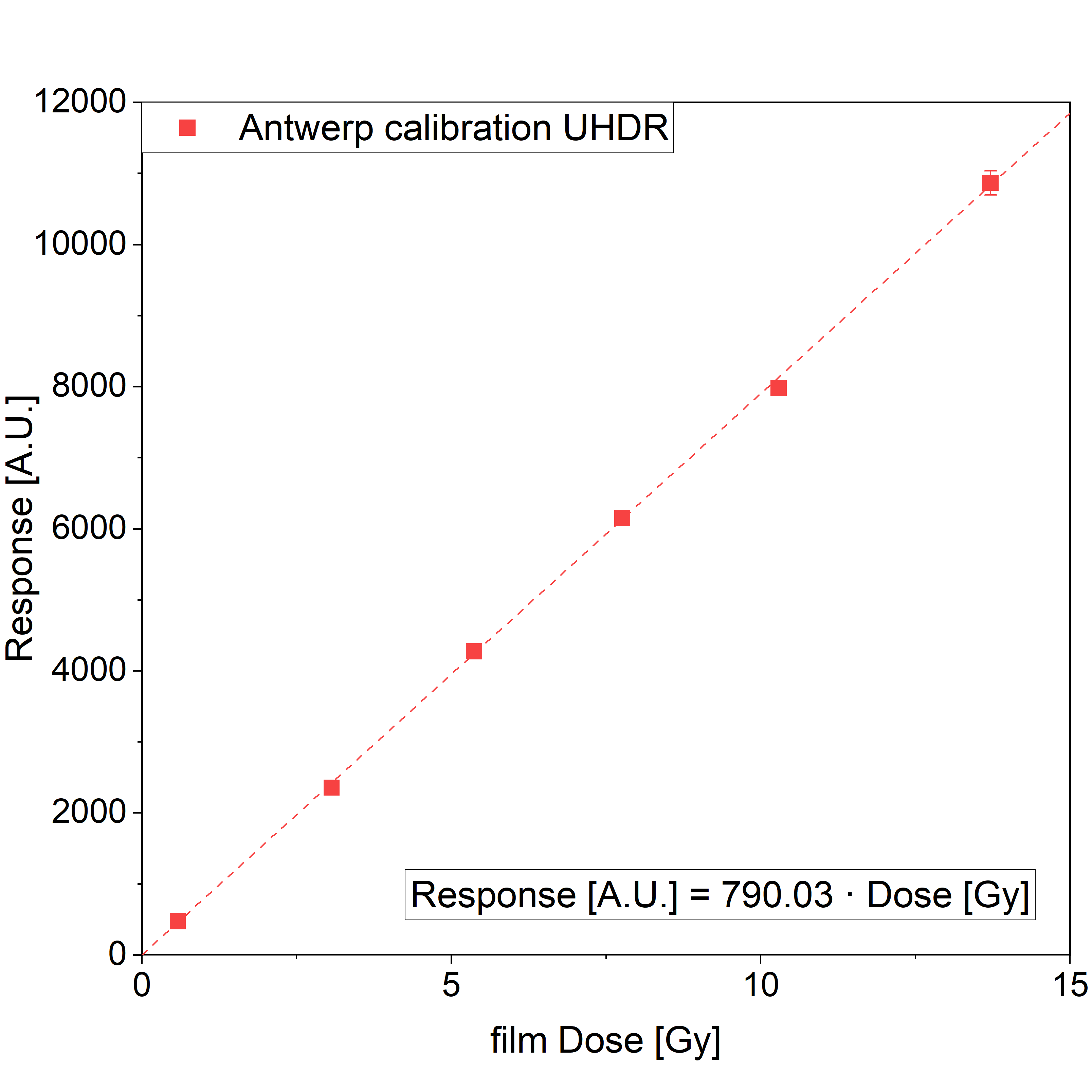}
    \caption{The dose response curve from the ImageDosis system and the radiochromic film dose with increasing number of pulses for the measurement campaign in the 9 MeV Antwerp beam.}
    \label{fig: DoseResponse}
\end{figure}

Figure~\ref{fig: DoseResponse} shows the dose response curve for doses ranging between 0 and 13 Gy in Antwerp. The average dose rate for these irradiations was 30.0 Gy/s.  The dashed lines denote the linear fit (R$^2 >$ 0.999) that determines the calibration factor, given by the formula in the graph. In the following sections, this calibration factor will be used to report the normalized response, defined as the response divided by the calibration factor and the radiochromic film dose. 
\par
\begin{figure}[!ht]
    \centering
    \includegraphics[width=\textwidth]{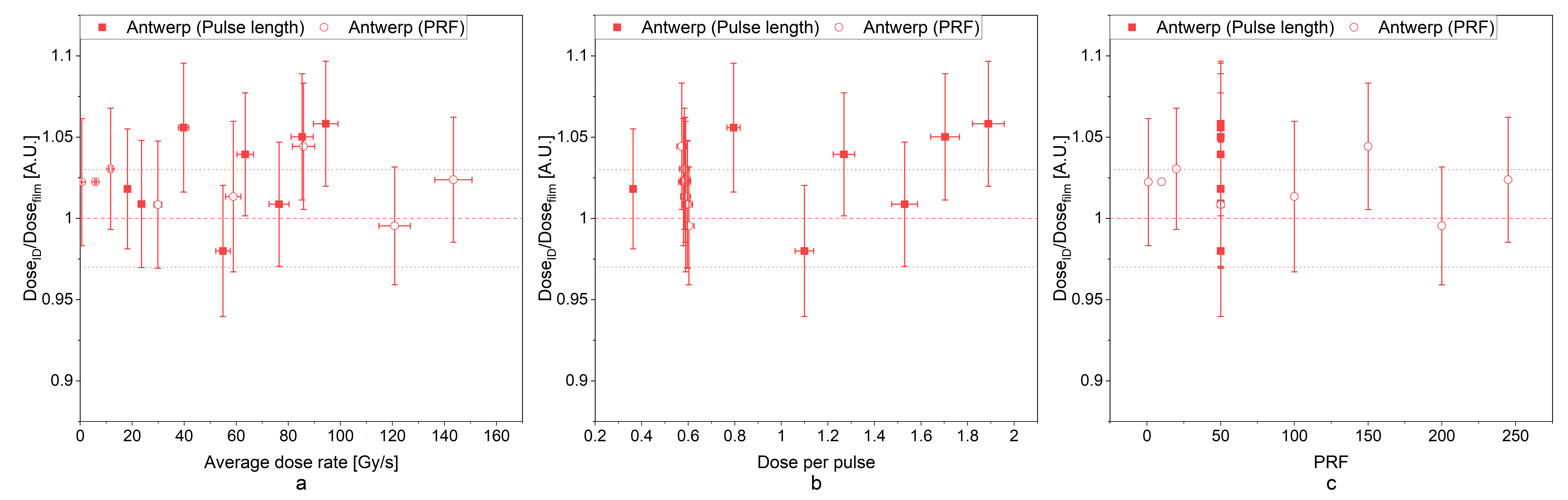}
    \caption{The ratio of the response from the ImageDosis system and the radiochromic film dose with varying (a) average dose rate, (b) dose per pulse and (c) PRF, for the measurement campaign in the 9 MeV Antwerp beam (red) and in the 5 MeV Orsay beam (gray).}
    \label{fig: allDoseRatedef}
\end{figure}

The normalized response with varying average dose rate, dose per pulse and PRF is shown in Figure~\ref{fig: allDoseRatedef}. The circles and squares denote the irradiations where, respectively, the pulse length and PRF were varied. The dashed line represents unity and the dotted lines represent 3\% uncertainty. Five data points fall outside the uncertainty band, all overestimating the dose. No trend is observed. The variation of the normalized dose with variation of the pulse length and pulse repetition frequency is 2.7\% and 1.4\% respectively.

\par
The response with energy was tested by variation of the nominal beam energy. The normalized response for the 7 and 9 MeV irradiations were 1.04 and 1.01, respectively, resulting in a difference of 3.4\%.

\par
The YAG- based ImageDosis system did not detect 2.3\% (16/679) of the pulses due to the absence of triggering. It showed to be capable to discriminate 2 subsequent pulses for all combinations of irradiation settings. 
\par
\begin{figure}[!ht]
    \centering
    \includegraphics[width = .7\textwidth]{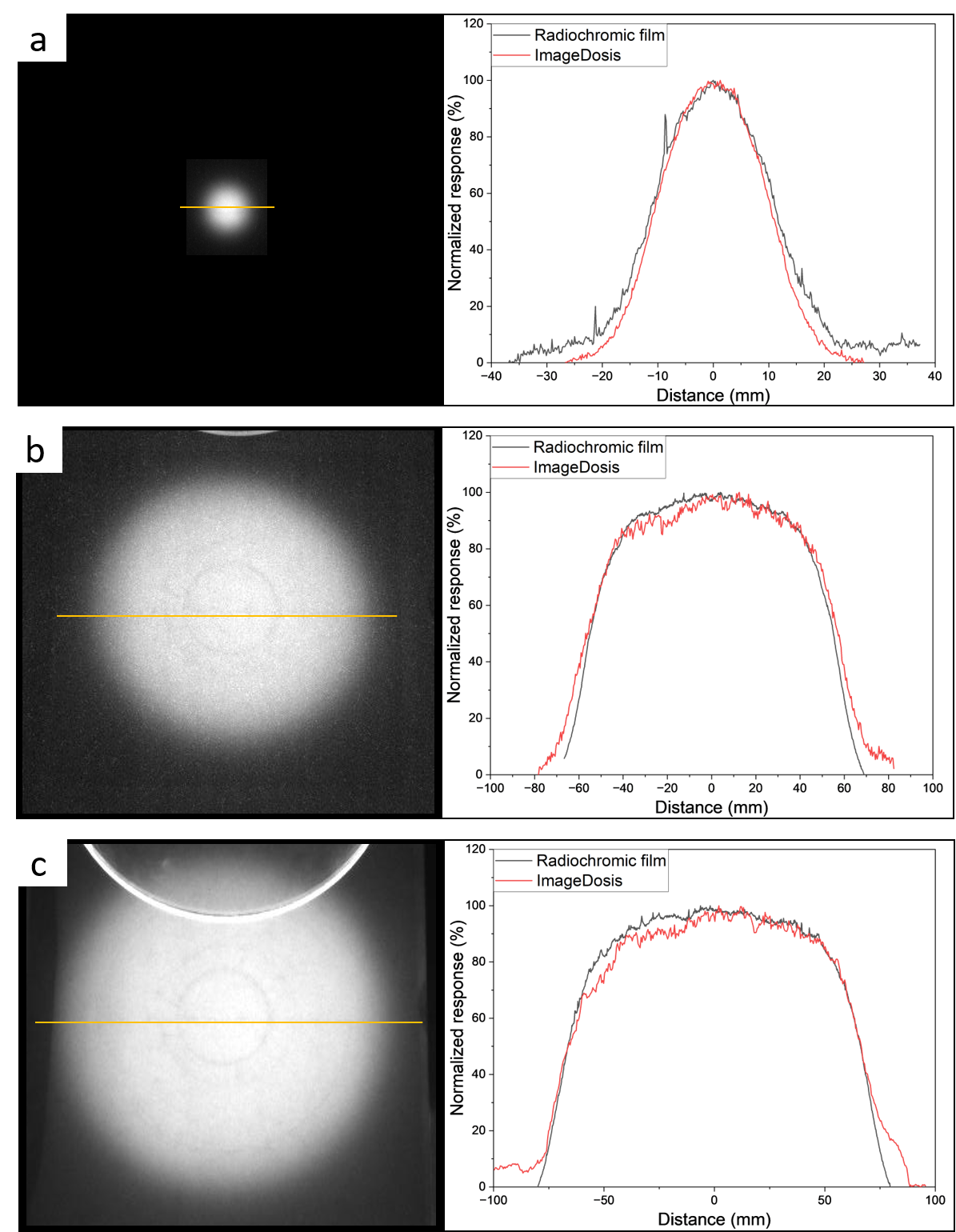}
    \caption{An ImageDosis frame with corresponding profile for a circular applicator with a diameter of 18 (a), 100 (b) and 120 mm (c). The yellow overlay on the ImageDosis frame shows the position of the profile.}
    \label{fig: 2D}
\end{figure}

The vertical profiles obtained with the ImageDosis system and radiochromic film for circular applicators with nominal diameters of 18 mm, 100 mm and 120 mm can be appreciated in Figure~\ref{fig: 2D}. The ImageDosis frames were rotated such that the profile and its position (shown in yellow on the frame) are aligned. The penumbra and FWHM measured by the ImageDosis system and the radiochromic films for all applicators are given in Table~\ref{tab: 2D characteristics}. For the 100 mm and 120 mm applicators the upper part (left part in the figure) of the ImageDosis profile shows a reduced output. The 18 mm applicator ImageDosis profile shows a slightly decreased output compared to the radiochromic film at the lower part of the penumbra. The 100 mm and 120 mm applicator ImageDosis profile show the opposite.  

\begin{table}[!ht]
    \centering
\caption{The penumbra (20\%-80\% width, averaged left and right) and FWHM for the 18 mm, 100 mm and 120 mm applicators, measured by radiochromic film and the ImageDosis system.}
\label{tab: 2D characteristics}
    \begin{tabular}{|c|c|c|c|c|c|c|}
        \hline
         \multirow{3}{*}{Applicator size}&  \multicolumn{3}{c|}{FWHM [mm]}& \multicolumn{3}{c|}{Penumbra [mm]}\\
         \cline{2-7}
        & radiochromic & \multirow{2}{*}{ImageDosis}& \multirow{2}{*}{$\Delta$ [\%]}& radiochromic  &\multirow{2}{*}{ImageDosis}& \multirow{2}{*}{$\Delta$ [\%]}\\
        &film&&&film&&\\
        \hline
         \multicolumn{1}{|c|}{18 mm}&  23.0 $\pm$ 0.2&22.2 $\pm$ 0.2 & 3.6 &9.5 $\pm$ 0.2&8.8 $\pm$ 0.2& 7.8\\
         \hline
         \multicolumn{1}{|c|}{100 mm}& 109.2 $\pm$ 0.2 & 113.8 $\pm$ 0.4 & 3.6 & 17.8 $\pm$ 0.2 & 20.5 $\pm$ 0.4 & 13.8\\
         \hline
         \multicolumn{1}{|c|}{120 mm}&  132.4 $\pm$ 0.2&132.4 $\pm$ 0.4 & 0.05 &18.8 $\pm$ 0.2&25.2 $\pm$ 0.6 & 29.0\\
         \hline
    \end{tabular}

\end{table}

\subsection{Validation}
The results obtained in Antwerp were validated, and the transferability of the ImageDosis system was investigated by a measurement campaign at Institut Curie in Orsay, where the another unit of the ElectronFlash is located. 
\par
\begin{figure}
    \centering
    \includegraphics[width=\textwidth]{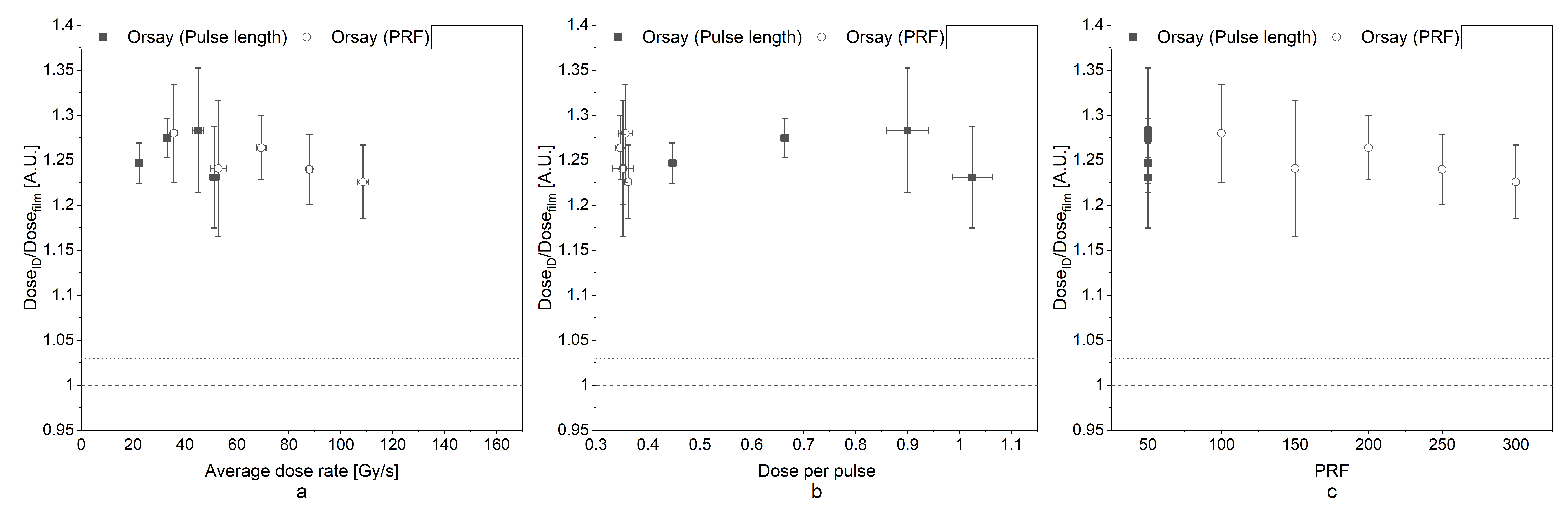}
    \caption{The ratio of the response from the ImageDosis system and the radiochromic film dose with varying (a) average dose rate, (b) dose per pulse and (c) PRF, for the measurement campaign in the 9 MeV Antwerp beam (red) and in the 5 MeV Orsay beam (gray).}
    \label{fig: OrsayDR}
\end{figure}

The normalized response with varying average dose rate, dose per pulse and PRF is shown in Figure~\ref{fig: OrsayDR}. The circles and squares denote the irradiations where, respectively, the pulse length and PRF were varied. The dashed line represents unity and the dotted lines represent 3\% uncertainty. An offset is observed, where all data points are spread around a normalized response of 1.24. The variation of the normalized response with varying pulse length and PRF is comparable with the Antwerp data, being 1.9\% and 1.7\%, respectively.
\par
The normalized responses for 5 and 7 MeV irradiations in Orsay, were 1.19 and 1.20, respectively, resulting in a 0.5\% difference. 
\par
In Orsay, the ImageDosis system did not detect 3.2\% (23/710) of the pulses. It was capable of discriminating 2 subsequent pulses for irradiations with PRF $<$ 300 Hz. After exclusion of the corrected pulses, a Pearson correlation coefficient of 0.68 was found when comparing the individual pulses measured by the ImageDosis with the ones measured by the point scintillating system for the data from the "response with dose" experiment. This is represented in Figure~\ref{fig: corr} (supplementary section).

\section{Discussion}
In order to appreciate the FLASH effect in preclinical experiments, 2D real time dosimetry with a temporal resolution in the µs range will be an added value. The ImageDosis system, with Ce-doped YAG as scintillating material, was investigated as a candidate to fulfill that need. Its response with dose, average dose rate, dose per pulse, PRF and energy were investigated for UHDR electron beams, and the temporal and 2D characteristics were touched upon. 
\par
The ImageDosis showed to scale linearly with dose for doses up to 13 Gy. The linear behavior is expected to remain valid for higher doses. In theory, the limiting factors are the instantaneous dose rate and the dynamic range of the camera. The calibration factor was extracted from the Antwerp data, but showed not to be transferable to another center. The normalized response from the system for the Orsay data was 24\% higher compared to the one measured in Antwerp. A minor contribution to this difference can be explained by the setup uncertainty since all parts of the ImageDosis system are positioned individually. However, we attribute the major contribution to the phantom material. In Antwerp RW3 was used, whereas in Orsay PMMA was used. PMMA is, in contrast to RW3, a transparent material, allowing Cherenkov radiation, generated in the phantom, to contribute to the signal, resulting in an increased response as is observed in the data. Where the Cherenkov radiation has proven to be useful for dosimetry~\cite{lee2016measurement,darafsheh2018characterization,rahman2021spatial}, it is an unwanted effect in this study. In point scintillators, the Cherenkov contribution can be accounted for via a spectral analysis~\cite{therriault2013nature}, however, for this 2D application a wavelength filter transmitting the luminescent light and blocking the Cherenkov light should be sufficient. 
\par
The response of this YAG-sheet based ImageDosis system can be considered independent of the average dose rate, dose per pulse and PRF, up to at least 140 Gy/s, 1.8 Gy/pulse and 300 Hz, respectively. Most data points from the Antwerp measurement campaign, fall within 3\% from unity, the variation between the data points is within 2.7\%, all error bars overlap and no trends are observed, irrespective of which parameter is varied. \newline
Leaving the increased response aside, the validation data from Orsay supports this conclusion. The variation between the data points is within 1.7\% and all error bars overlap. A decreasing trend with PRF can be observed, however, this is minor and within the uncertainty of the data points. Further optimization, decreasing the uncertainties, is needed to investigate a potential PRF dependence. \newline
The majority of preclinical experiments and works on future clinical applications, report a PRF between 100 and 200 Hz and a dose per pulse around 1.5 Gy~\cite{bourhis2019clinical,bourhis2019treatment,vozenin2022towards}. This work shows that the ImageDosis system is average dose rate, dose per pulse and PRF independent in the range that is currently considered relevant for FLASH-RT. However, the field is rapidly evolving and a few studies report the FLASH effect for PRF up to 300 Hz and dose per pulse up to 12 Gy~\cite{vozenin2019advantage,montay2019long}. Therefore, this initial characterization needs to be extended and investigation of the ImageDosis system with increased dose per pulse, PRF and instantaneous dose rate is needed. In addition, before relying solely on the ImageDosis system in a (pre)clinical setting, further optimization is recommended to decrease the uncertainty. 
\par
Due to the increased response for the Orsay measurement campaign, no direct comparison could be performed between the normalized responses to beams with energies of 5, 7 and 9 MeV. Therefore, only a pairwise comparison between the 5 and 7 MeV, and the 7 and 9 MeV beams could be performed. From this, no energy dependence was observed for the YAG-coating, as the normalized response for both the 5 and 7 MeV, and the 7 and 9 MeV irradiations fell within their respective uncertainties. In an attempt to compare the normalized response for the 3 energies, the normalized responses from the Orsay measurement campaign were corrected such that the 7 MeV responses matched. Hereafter, the variation with energy is 1.8\%, supporting the energy independence statement. In contrast with ionization chambers, which require an energy dependent beam quality correction factor~\cite{musolino2001absorbed}, the YAG-based ImageDosis system does not require an additional energy correction factor, as such eliminating an additional source of uncertainty. 
\par
One advantage of this camera-scintillating sheet system is the high temporal resolution, which is crucial for real time UHDR dosimetry. The fact that about 3\% of the pulses were not detected is solely due to the lack of triggering during these measurement campaigns. Without triggering, a pulse can be delivered at the time the shutter of the camera is closed between frames, preventing the system from detecting this pulse. In fact, the undetected pulses prove that the decay time of the scintillating sheet is sufficiently small for use in UHDR beams with a PRF up to 245 Hz. If the frame rate of the camera is sufficiently high (i.e. $> 2$ times the PRF), this decay time is the limiting factor, that determines the temporal resolution. The Ce-doped YAG has a decay time of 70 ns and is, therefore, suited for dosimetry of irradiations with very high PRF. 
\newline
The correlation between the pulse amplitudes detected by the YAG-based ImageDosis system and the point scintillator, even within a single irradiation, indicates that both systems are capable of measuring small output variations of the system. Since both systems are research prototypes, and the point scintillator was used for pulse discrimination rather than dosimetry, the lower limit of such variations that can be detected could not be investigated. Nevertheless, the correlation is expected to increase when both systems are optimized and irradiated under reference conditions. 
\par
Another benefit of the ImageDosis systems is its 2D character. For the 18 mm the ImageDosis system and radiochromic film show almost perfect overlap within the field. However, in the lower part of the penumbra, at the edge of the field, the dose measured by the ImageDosis system is lower than the one measured by radiochromic film. It is known that radiochromic film overestimates the dose for low energy photons~\cite{ju2002film, cheung2006measurement} and, therefore, it is plausible that the observed discrepancy is caused by the deviating film measurement. Further investigation is needed to verify this hypothesis. Also to be investigated is the decreased output on the upper part (left in Figure~\ref{fig: 2D}) of the field for the 100 mm and 120 mm applicators. It is plausible that this decrease is in fact due to the high transparency of the scintillating sheet. If the reflection of the phantom is not homogeneous, this may affect the measurement. Another possibility is an inhomogeneity of the coating. The latter is, however, not to be expected as the sheet was coated in a production environment where high homogeneity is required. Also, the overresponse outside the field requires further investigation.  \newline
A plausible reason for the slightly altered penumbras, is the limited processing of the ImageDosis image. As the camera is capturing the scintillating sheet under an angle, an affine transformation is needed to generate a perpendicular view on the sheet. This can easily be performed when an additional frame with a checkerboard pattern is recorded~\cite{rahman2020characterization}. The checkerboard image has the additional benefit that the pixel size can be determined more accurately, which could improve the overlap between the profiles. After updating the processing pipeline, also horizontal profiles can be extracted and compared.

\subsection{Further perspectives}
The main limiting factors to this research are the difference in phantom material between the characterization and validation; the lack of triggering; and the use of images which are not corrected for angled acquisition. New data is being acquired, using an RW3 phantom, with triggered camera control to investigate the full potential of the YAG-based ImageDosis system. This has the additional benefit of strongly reducing the required storage, making the frame rate setting limitation irrelevant. This is expected to allow pulse discrimination for irradiations with higher PRF. Also, the affine transformation is being automated and implemented in the pipeline.  
Finally, the direct comparison between the point scintillator and the ImageDosis system to determine its limits, and standardization of the setup are future works.

\section{Conclusions}
In this study, the ImageDosis system with a camera with high frame rate and Ce-doped YAG scintillating sheet was characterized and validated as 2D dosimeter for UHDR electron dosimetry. It showed to be linear with dose for doses up to at least 13 Gy. The system showed independence against average dose rate up to 140 Gy/s, dose per pulse up to 2 Gy and PRF up to 300 Hz on two different irradiation machines. Also, no energy dependence was observed. One of the major benefits of this system is the real time dose assessment. It showed to be able to measure the output of individual pulses up to pulse repetition frequencies of 245 Hz, which can potentially be improved with the current implementation of the camera triggering. The 2-dimensional characteristics showed sub-optimal profiles compared with film, which require further investigation after updating the processing pipeline.

\section{Acknowledgments}
This work was supported by VLAIO via the Flanders.HealthTech call [HBC.2021.0946] and by FWO via the Tournesol grant [VS00823N]. SCK CEN and Iridium Network are also non-funded collaborators to the 18HLT04 UHDpulse project which received funding from the EMPIR programme. 

\newpage
\printbibliography
\newpage
\appendix
\setcounter{table}{0}
\renewcommand{\thetable}{A\arabic{table}}
\setcounter{figure}{0}
\renewcommand{\thefigure}{A\arabic{figure}}

\section{Supplementary}

\begin{table}[!ht]
    \caption{The camera specifications and settings.}
    \centering
    \begin{tabular}{|c|c| }
    \hline
    \textbf{Parameter/setting} & \textbf{C-BLUE ONE}\\
    \hline
    Physical dimensions WxLxH [cm$^3$] &  6.4x7.6x15.4\\
    \hline
    Sensor type &  CMOS \\
    \hline
    Pixel pitch [µm]  & 9\\
    \hline
    Resolution &  816x624\\
    \hline
    Quantum efficiency at 650 nm [\%] & 70\\
    \hline
    Maximum frame rate [fps] &  1594\\
    \hline
    Used frame rate [fps] &  500.3\\
    \hline
    Integration time [µs] &  1945.1 \\
    \hline
    Calculated dead time [µs] & 53.5\\
    \hline
    Gain [dB]& 24\\
    \hline
    \end{tabular}
    \label{tab: camera settings}
\end{table}

\begin{table}[!ht]
    \centering
    \caption{The beam parameters used for all experiments. The average dose rate, dose per pulse and instantaneous dose rate values were experimentally obtained with radiochromic film.}
    \resizebox{\textwidth}{!}{ 
    \begin{tabular}{|c|c|c|c|c|c|c|c|c|c|c| }
    \hline
         \multirow{2}{*}{Experiment} & \multirow{2}{*}{Facility} &  Energy & Field size & Pulse repetition & pulse length  & source surface & Number of  &  Average dose & Dose per & Instantaneous dose \\
         & &[MeV] & $\diameter$[cm] & frequency [Hz] & [$\mu$s] & distance [cm] &pulses&  rate [Gy/s] & pulse [Gy] & rate [MGy/s]\\
         \hline
         \multirow{11}{*}{Response with Dose}&  Antwerp & 9 & 12 & 50 & 1 & 105.0 & 1 & 38.06 & 0.76 & 0.76   \\
          \cline{2-11}
          & Antwerp & 9 & 12 & 50 & 1 & 105.0 & 5 & 30.21 & 0.60 & 0.60 \\
          \cline{2-11}
          & Antwerp & 9 & 12 & 50 & 1 & 105.0 & 9 & 28.37 & 0.57 & 0.57   \\
          \cline{2-11}
          & Antwerp & 9 & 12 & 50 & 1 & 105.0 & 13 & 28.04 & 0.56 & 0.56   \\
          \cline{2-11}
          & Antwerp & 9 & 12 & 50 &  1 & 105.0 & 17 & 28.01 & 0.56 & 0.56   \\
          \cline{2-11}
          & Antwerp & 9 & 12 & 50 & 1 & 105.0 & 23 & 27.57 & 0.55 & 0.55   \\
          \cline{2-11}
          & Orsay & 5 & 12 & 50 & 1 & 105.0 & 1 & 26.53 & 0.53 & 0.53   \\
          \cline{2-11}
          & Orsay & 5 & 12 & 50 & 1 & 105.0 & 5 & 19.99 & 0.40 & 0.40   \\
          \cline{2-11}
          & Orsay & 5 & 12 & 50 & 1 & 105.0 & 10 & 18.38 & 0.37 & 0.37   \\
          \cline{2-11}
          & Orsay & 5 & 12 & 50 & 1 & 105.0 & 15 & 18.21 & 0.36 & 0.36  \\
          \cline{2-11}
          & Orsay & 5 & 12 & 50 & 1 & 105.0 & 20 & 18.33 & 0.37 & 0.37   \\
          \hline
          \multirow{12}{*}{PRF} & Antwerp & 9 & 12 & 1 & 1 & 105.0 & 9 & 0.56 & 0.56 & 0.56 \\
          \cline{2-11}
          & Antwerp & 9 & 12 & 10 & 1 & 105.0 & 9 & 5.66 & 0.57 & 0.57 \\
          \cline{2-11}
          & Antwerp & 9 & 12 & 20 & 1 & 105.0 & 9 & 11.19 & 0.56 & 0.56 \\
          \cline{2-11}
          & Antwerp & 9 & 12 & 100 & 1 & 105.0 & 9 & 56.25 & 0.56 & 0.56 \\
          \cline{2-11}
          & Antwerp & 9 & 12 & 150 & 1 & 105.0 & 9 & 82.52 & 0.55 & 0.55 \\
          \cline{2-11}
          & Antwerp & 9 & 12 & 200 & 1 & 105.0 & 9 & 114.97 & 0.57 & 0.57 \\
          \cline{2-11}
          & Antwerp & 9 & 12 & 245 & 1 & 105.0 & 9 & 137.90 & 0.56 & 0.56 \\
          \cline{2-11}
          & Orsay & 5 & 12 & 100 & 1 & 105.0 & 20 & 34.92 & 0.35 & 0.35   \\
           \cline{2-11}
          & Orsay & 5 & 12 & 150 & 1 & 105.0 & 20 & 51.82 & 0.35 & 0.35   \\
           \cline{2-11}
          & Orsay & 5 & 12 & 200 & 1 & 105.0 & 20 & 68.02 & 0.34 & 0.34   \\
           \cline{2-11}
          & Orsay & 5 & 12 & 250 & 1 & 105.0 & 20 & 86.11 & 0.34 & 0.34   \\
           \cline{2-11}
          & Orsay & 5 & 12 & 300 & 1 & 105.0 & 20 & 106.37 & 0.35 & 0.35   \\
          \hline
          \multirow{12}{*}{Pulse length} & Antwerp & 9 & 12 & 50 & 0.5 & 105.0 & 16 & 17.36 & 0.35 & 0.69   \\
          \cline{2-11}
          & Antwerp & 9 & 12 & 50 & 1.5 & 105.0 & 6 & 38.20 & 0.76 & 0.51   \\
          \cline{2-11}
          & Antwerp & 9 & 12 & 50 & 2 & 105.0 & 4 & 53.16 & 1.06 & 0.53   \\
          \cline{2-11}
          & Antwerp & 9 & 12 & 50 & 2.5 & 105.0 & 3 & 62.06 & 1.24 & 0.50   \\
          \cline{2-11}
          & Antwerp & 9 & 12 & 50 & 3 & 105.0 & 3 & 73.75 & 1.47 & 0.49   \\
          \cline{2-11}
          & Antwerp & 9 & 12 & 50 & 3.5 & 105.0 & 2 & 83.76 & 1.68 & 0.48   \\
          \cline{2-11}
          & Antwerp & 9 & 12 & 50 & 4 & 105.0 & 2 & 92.42 & 1.85 & 0.46   \\
          \cline{2-11}
          & Orsay & 5 & 12 & 50 & 0.69 & 105.0 & 7 & 10.90 & 0.22 & 0.32   \\
          \cline{2-11}
          & Orsay & 5 & 12 & 50 & 1.47 & 105.0 & 7 & 21.77 & 0.44 & 0.30   \\
          \cline{2-11}
          & Orsay & 5 & 12 & 50 & 2.36 & 105.0 & 7 & 32.41 & 0.65 & 0.27   \\
          \cline{2-11}
          & Orsay & 5 & 12 & 50 & 3.19 & 105.0 & 7 & 44.00 & 0.88 & 0.28   \\
          \cline{2-11}
          & Orsay & 5 & 12 & 50 & 3.98 & 105.0 & 7 & 50.24 & 1.00 & 0.25   \\
          \hline
          \multirow{4}{*}{Energy} & Antwerp & 9 & 12 & 50 & 1 & 105.0 & 9 & 28.37 & 0.57 & 0.57  \\
          \cline{2-11}
          & Antwerp & 7 & 12 & 50 & 1 & 105.0 & 56 & 8.77 & 0.18 & 0.18   \\
          \cline{2-11}
          & Orsay & 7 & 12 & 50 & 1 & 105.0 & 20 & 37.97 & 0.76 & 0.76   \\
          \cline{2-11}
          & Orsay & 5 & 12 & 50 & 1 & 105.0 & 30 & 19.83 & 0.40 & 0.40   \\
          \hline
          \multirow{2}{*}{2D} & Antwerp & 9 & 12 & 10 & 4 & 105.0 & 6 & N.A. & N.A. & N.A.  \\
          \cline{2-11}
          & Antwerp & 9 & 1.8 & 10 & 4 & 105.0 & 1 & N.A. & N.A. & N.A.  \\
          \hline
    \end{tabular}}
    \label{tab: Beam parameters}
\end{table}

\begin{figure}[!ht]
    \centering
    \includegraphics[width = .7\textwidth]{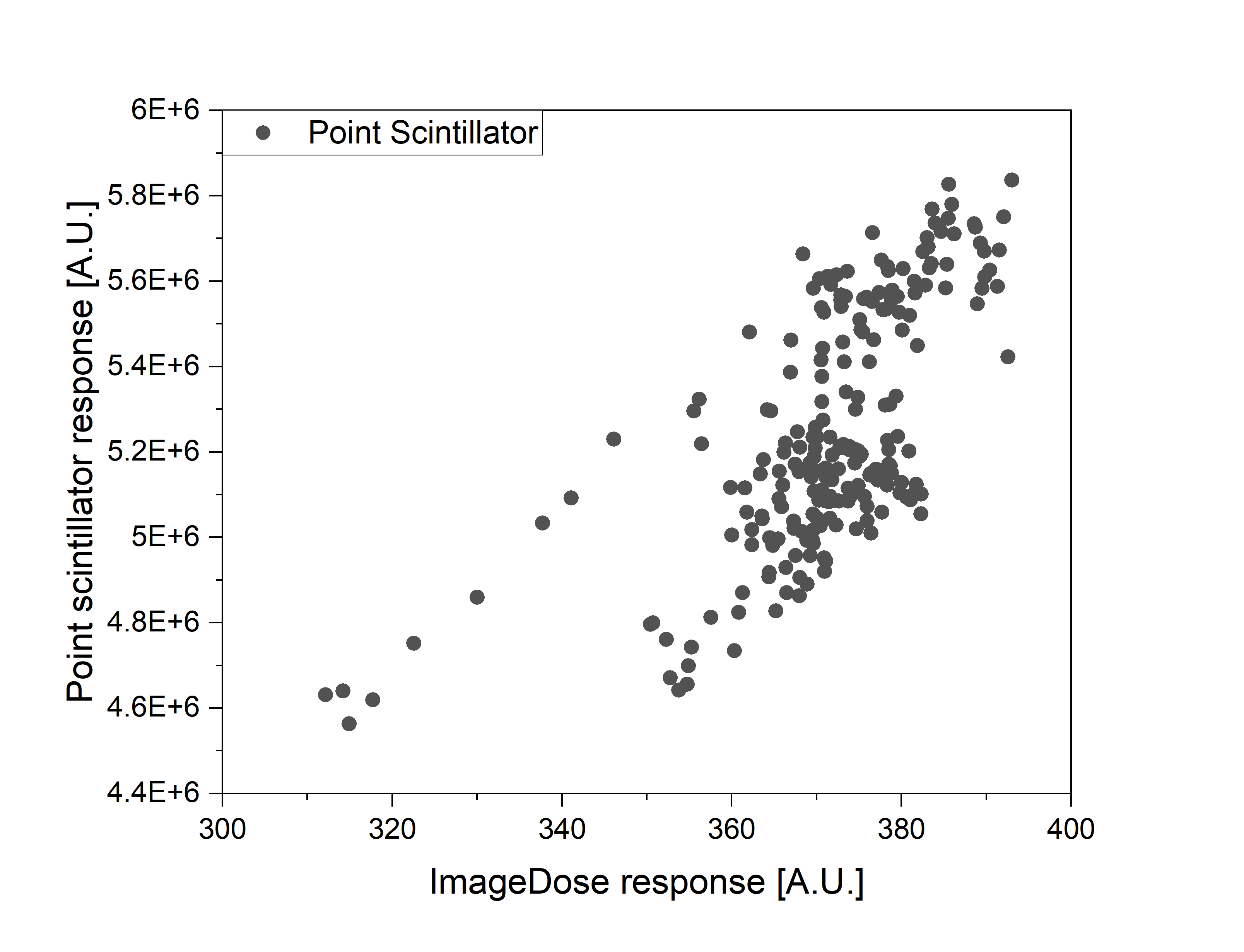}
    \caption{The point scintillator response against the ImageDosis response for each individual pulse from the response with dose experiment.}
    \label{fig: corr}
\end{figure}

\end{document}